\documentclass[a4paper,12pt,twoside]{article}
\usepackage[english]{babel}
\usepackage{rotating}
\usepackage{epsfig}
\usepackage[centertags]{amsmath}
\usepackage{array}
\usepackage{multirow}
\usepackage{supertabular}
\usepackage{dcolumn}
\usepackage{cite}
\usepackage{xspace}
\def\zero{{\scriptscriptstyle 0}}

\def\Z0{\ensuremath{Z^\zero}}

\def\SU2U1{{\rm SU}(2)\times{\rm U}(1)}

\def\meas{{\rm meas}}

\mathchardef\qsm=63
\mathchardef\pls=43
\mathchardef\mns=512
\mathchardef\plm=518
\mathchardef\eql=61
\mathchardef\smallleft=300
\mathchardef\smallright=301
\mathchardef\perslsh=47
\mathchardef\les=316
\mathchardef\gre=318
\mathchardef\leq=532
\mathchardef\grq=533
\chardef\usc=95
\chardef\til=126

\def\sqr#1#2#3{{\vcenter{\hrule height.#3ex\hbox{\vrule width.#2ex height#1ex
    \kern#1ex\vrule width.#3ex}\hrule height.#2ex}}}

\def\angleto{\vrule width.035em height2.1ex depth-.56ex\unskip\kern-.6ex\to}
\def\perchc#1{{\raise.4ex\hbox{$\mkern4mu#1{\it\perslsh}_
             {\mkern-5mu\scriptscriptstyle{{\rm o}\!{\rm o}}}^
             {\mkern-12.8mu\scriptscriptstyle{\rm o}}$}}}

\catcode`\@=11 
\def\parenbar{\mathpalette\p@renb@r}
\def\p@renb@r#1#2{\vbox{%
  \ifx#1\scriptscriptstyle \dimen@.7em\dimen@ii.2em\else
  \ifx#1\scriptstyle \dimen@.8em\dimen@ii.25em\else
  \dimen@1em\dimen@ii.4em\fi\fi \offinterlineskip
  \ialign{\hfill##\hfill\cr
    \vbox{\hrule width\dimen@ii}\cr
    \noalign{\vskip-.3ex}%
    \hbox to\dimen@{$\mathchar300\hfil\mathchar301$}\cr
    \noalign{\vskip-.3ex}%
    $#1#2$\cr}}}
\catcode`\@=12 

\newbox\struttbox
\setbox\struttbox=\hbox{\vrule height1.65ex depth.485ex width0pt}
\def\strutt{\relax\ifmmode\copy\struttbox\else\unhcopy\struttbox\fi}
\def\stru#1#2{\relax\ifmmode\hbox{\vrule height#1 depth#2 width0pt}
\else\vrule height#1 depth#2 width0pt\fi}

\def\ronum#1{\uppercase\expandafter{\romannumeral#1}}
\def\ronuml#1{\expandafter{\romannumeral#1}}

\DeclareMathAlphabet{\mathbf}{OT1}{cmr}{bx}{sl}

 {\end{list}}
 {\end{list}}
 {\end{list}}

\catcode`\@=11 
\newlength{\@fninsert}
\newlength{\@fnwidth}
\renewcommand{\@makefntext}[1]%
  {\noindent\makebox[\@fninsert][r]{\@makefnmark}\hfil%
  \parbox[t]{\@fnwidth}{#1}}
\catcode`\@=12 
\newlength{\localtextwidth}
\catcode`\@=11 
\newsavebox{\tmpbox}
\newlength{\@captionmargin}
\newlength{\@captionwidth}
\newlength{\@captionitemtextsep}
\renewcommand{\@makecaption}[2]%
  {%
   \vspace{10.pt}
   \setlength{\@captionwidth}{\localtextwidth}
   \addtolength{\@captionwidth}{-\@captionmargin}
   \sbox{\tmpbox}{{\bf #1:}{\it #2}}%
   \ifthenelse{\lengthtest{\wd\tmpbox > \@captionwidth}}%
   {\centerline{\parbox[t]{\@captionwidth}%
   {\tolerance=2000\normalsize%
    {\rm #1:}\hspace{\@captionitemtextsep}{\rm #2}}}}%
   {\centerline{{\bf #1:}\kern1.em{\it #2}}}}
\catcode`\@=12 
\catcode`\@=11 
\renewcommand\section{\@startsection{section}{1}{\z@}%
                                   {-3.5ex \@plus -1ex \@minus -.2ex}%
                                   {2.3ex \@plus.2ex}%
                                   {\normalfont\Large\bfseries}}
\renewcommand\subsection{\@startsection{subsection}{2}{\z@}%
                                   {-3.25ex\@plus -1ex \@minus -.2ex}%
                                   {1.5ex \@plus .2ex}%
                                   {\normalfont\large\bfseries}}
\renewcommand\subsubsection{\@startsection{subsubsection}{3}{\z@}%
                                   {-3.25ex\@plus -1ex \@minus -.2ex}%
                                   {1.5ex \@plus .2ex}%
                                   {\normalfont\large\bfseries}}
\renewcommand\paragraph{\@startsection{paragraph}{4}{\z@}%
                                   {3.25ex \@plus1ex \@minus.2ex}%
                                   {1.2ex \@plus .2ex}%
                                   {\normalfont\normalsize\bfseries}}
\catcode`\@=12 

\newsavebox{\sesbox}
\newlength{\seslen}

\newcommand{\kt}{\mbox{$<\!\!k_T\!\!>\,$}}
\newcommand{\ktintr}{\mbox{$<\!\!k_T^{\,\mathrm{intr}}\!\!>\,$}}
\newcommand{\pprms}{\mbox{$<\!\!p_\perp^{\,\mathrm{rms}}\!\!>\,$}}
\newcommand{\ks}{\mbox{$k_0$}}
\newcommand{\Qt}{\mbox{$<\!\!Q_T\!\!>\,$}}

\newcommand{\jet}{{\,\mathrm{jet}}}
\renewcommand{\meas}{{\,\mathrm{meas}}}
\newcommand{\yJB}{y_{\mathrm{JB}}}
\newcommand{\pt}{\mbox{$p_T$}}
\newcommand{\Et}{\mbox{$E_T$}}
\begin{document}
\selectlanguage{english}

\title{
\bf\LARGE Study of the effective transverse momentum of partons in the 
proton using prompt photons in  photoproduction at HERA\\
\vspace{-4.8cm}
\begin{flushleft}
\tt\normalsize DESY 01-043\\
Mar.\ 2001\\[5cm]
\end{flushleft}
}                                                       
                    
\author{ZEUS Collaboration}
\date{ }

\maketitle

\vfill

\centerline{\bf Abstract}
\vskip4.mm
\begin{center}
  \begin{minipage}{15.cm}
    \noindent
   
    The photoproduction of prompt photons, together with an
accompanying jet, has been measured with the ZEUS detector at HERA
using an integrated luminosity of 38.6 pb$^{-1}$.  A study of the
effective transverse momentum, \kt, of partons in the proton, as
modelled within the framework of the PYTHIA Monte Carlo, gives a value
of $ \kt = 1.69\pm0.18\; ^{+0.18}_{-0.20} $ GeV for the $\gamma p$
centre-of-mass energy range $134 < W < 251$ GeV.  This result is in
agreement with the previously observed trend in hadron-hadron
scattering  for \kt\ to rise with interaction energy.
\end{minipage}
\end{center}

\vfill
\thispagestyle{empty}

\newpage
\parindent0.cm                                                                                     
\parskip0.3cm plus0.05cm minus0.05cm                                                               
\def\3{\ss}                                                                                        
\newcommand{\address}{ }                                                                           
\newcommand{\author}{ }                                                                          
\pagenumbering{Roman}                                                                              
                                                   %
\begin{center}                                                                                     
{                      \Large  The ZEUS Collaboration              }                               
\end{center}                                                                                       
  S.~Chekanov,                                                                                     
  M.~Derrick,                                                                                      
  D.~Krakauer,                                                                                     
  S.~Magill,                                                                                       
  B.~Musgrave,                                                                                     
  A.~Pellegrino,                                                                                   
  J.~Repond,                                                                                       
  R.~Stanek,                                                                                       
  R.~Yoshida\\                                                                                     
 {\it Argonne National Laboratory, Argonne, IL, USA}~$^{p}$                                        
\par \filbreak                                                                                     
  M.C.K.~Mattingly \\                                                                              
 {\it Andrews University, Berrien Springs, MI, USA}
% ...
\newpage                                                                                           
$^{\    1}$ now visiting scientist at DESY \\                                                      
% ...
                                                           %
                                                           %
\newpage   
                                                           %
                                                           %
\begin{tabular}[h]{rp{14cm}}                                                                       
$^{a}$ &  supported by the Natural Sciences and Engineering Research                               
          Council of Canada (NSERC)  \\                                                            
$^{b}$ &  supported by the FCAR of Qu\'ebec, Canada  \\                                            
$^{c}$ &  supported by the German Federal Ministry for Education and                               
          Science, Research and Technology (BMBF), under contract                                  
          numbers 057BN19P, 057FR19P, 057HH19P, 057HH29P, 057SI75I \\                              
$^{d}$ &  supported by the MINERVA Gesellschaft f\"ur Forschung GmbH, the                          
          Israel Science Foundation, the U.S.-Israel Binational Science                            
          Foundation, the Israel Ministry of Science and the Benozyio Center                       
          for High Energy Physics\\                                                                
$^{e}$ &  supported by the German-Israeli Foundation, the Israel Science                           
          Foundation, and by the Israel Ministry of Science \\                                     
$^{f}$ &  supported by the Italian National Institute for Nuclear Physics                          
          (INFN) \\                                                                                
$^{g}$ &  supported by the Japanese Ministry of Education, Science and                             
          Culture (the Monbusho) and its grants for Scientific Research \\                         
$^{h}$ &  supported by the Korean Ministry of Education and Korea Science                          
          and Engineering Foundation  \\                                                           
$^{i}$ &  supported by the Netherlands Foundation for Research on                                  
          Matter (FOM) \\                                                                          
$^{j}$ &  supported by the Polish State Committee for Scientific Research,                         
          grant No. 111/E-356/SPUB-M/DESY/P-03/DZ 3001/2000,                                       
          620/E-77/SPUB-M/DESY/P-03/DZ 247/2000, and by the German Federal                         
          Ministry of Education and Science, Research and Technology (BMBF)\\                      
$^{l}$ &  partially supported by the German Federal Ministry for                                   
          Education and Science, Research and Technology (BMBF)  \\                                
$^{m}$ &  supported by the Fund for Fundamental Research of Russian Ministry                       
          for Science and Edu\-cation and by the German Federal Ministry for                       
          Education and Science, Research and Technology (BMBF) \\                                 
$^{n}$ &  supported by the Spanish Ministry of Education                                           
          and Science through funds provided by CICYT \\                                           
$^{o}$ &  supported by the Particle Physics and                                                    
          Astronomy Research Council \\                                                            
$^{p}$ &  supported by the US Department of Energy \\                                              
$^{q}$ &  supported by the US National Science Foundation                                          
\end{tabular}                                                                                      
                                                           %
                                                           %

\newpage
\pagenumbering{arabic}                                                                              
\setcounter{page}{1}
\parindent3pt

\section{Introduction}

The study of hard final-state photons in high-energy collisions is a
powerful tool for the investigation of parton dynamics and hadron
structure.  Photons of this kind (`prompt photons') can emerge as a
primary product of hard parton-scattering processes without the
hadronisation by which outgoing quarks and gluons form
observed jets.  In this way, they provide information about the
underlying parton processes that is relatively free from hadronisation
uncertainties.

In a recent analysis \cite{Z2}, ZEUS presented inclusive measurements
of prompt photon cross sections in photoproduction at HERA.  The
present paper describes a further study of such processes in which a
hadron jet is also measured.  The presence of the jet allows the
underlying QCD process in the $\gamma p$ interaction to be identified
more clearly, thus assisting the study of its dynamics.  This work is
motivated by the observation in a number of previous
experiments~\cite{CDF,D01,E706,D0}, summarised in recent
reviews~\cite{summary, wgreport}, that the inclusive production of
prompt photons with low transverse energy in hadron-proton and
hadron-nucleus reactions is unexpectedly large.  One possible
explanation is that the partons in the proton may effectively have a
considerably higher mean intrinsic transverse momentum, \kt, than the
traditionally assumed value of a few hundred MeV.  Measurements by
CDF~\cite{CDF} are consistent with a \kt\ of 3.5 GeV, while in
measurements at lower energies by E706~\cite{E706}, a value of 1.2 GeV
is suggested.  Recently published results from D0~\cite{D0} are
consistent with those of CDF.

The quantity \kt\ has been measured directly from the kinematics of
lepton or photon pairs that emerge from a hard interaction, but may
also be measured indirectly by making use of a theoretical framework
given by a next-to-leading-order (NLO) QCD calculation or a
leading-order Monte Carlo generator such as PYTHIA. The magnitude of
\kt\ has been taken to reflect the confinement of quarks and the known
size of the proton, with the assumption that these non-perturbative
effects may be combined in a straightforward way with a perturbative
calculation of the parton scattering.  However, it has more recently
been argued that when partons undergo hard scattering, the presence of
additional initial-state gluon radiation beyond NLO in QCD can
increase their effective \kt\ value and that this may be a major
contribution to the effects
observed~\cite{gluons,gluons2,durham1,Hoyer,Kimber,gluons3}. Within
PYTHIA, both of these contributions are allowed for: there is an
`intrinsic' component together with a parton-shower component, and
their combination is referred to here as the effective value of \kt.

In part of the measured pseudorapidity range, the ZEUS inclusive
prompt photon cross sections \cite{Z2} were found to be higher than
predicted.  However, Monte Carlo studies have indicated that
increasing \kt\ in the proton or photon is unable to account for this
discrepancy.  At the same time, the shape of the distribution in
transverse energy, \Et, can be well described by NLO theory, and the
overall normalisation, although slightly higher than predicted, is
insensitive to variations of \kt\, given the current experimental
statistical accuracy.  The aim of the present measurement is,
therefore, to determine by a more direct kinematic method whether the
partons in the proton possess high values of \kt\ in interactions with
a high-energy photon.  This is facilitated by the use of event samples
in which the `direct photoproduction' process dominates~\cite{Z1},
i.e.\ in which the entire incoming photon interacts with a quark in
the proton, thereby avoiding any additional contributions to \kt\ from the
resolved photon.  At leading order in photoproduction, the Compton
process $\gamma q \to \gamma q$ is the only direct prompt photon
process.

\section{Apparatus and Method}
The data were taken with the ZEUS detector at HERA, using an
integrated $e^+p$ luminosity of $38.6\pm0.6$ pb$^{-1}$.  The energies of
the incoming positron and proton were, respectively, $E_e = 27.5$ GeV
and $E_p = 820 $ GeV.  The apparatus and the details of the analysis
method are described in detail elsewhere~\cite{Z2,Z1}.  Of particular
relevance here are the compensating uranium-scintillator calorimeter
\cite{CAL} and the central tracking detector (CTD)~\cite{CTD}.  The
calorimeter provides almost hermetic coverage and has a relative energy
resolution, as measured in test beams, of $0.35/\sqrt{E}$ for hadronic
deposits and $0.18/\sqrt{E}$ for electromagnetic deposits, where $E$
is in GeV.  The CTD operates in a magnetic field of \( 1.43\,
\textrm{T} \) provided by a thin superconducting solenoid. The
transverse momentum resolution in the central rapidity region is \(
\sigma(\pt)/\pt =0.0058\,\pt \oplus 0.0065\oplus {0.0014/\pt } \) (\(
\pt \) in GeV).

Photons used in the present analysis were measured in the
electromagnetic section of the barrel region of the calorimeter, which
covers the polar angular range\footnote{ The ZEUS coordinate system is
a right-handed Cartesian system, with the $Z$ axis pointing in the
proton beam direction, referred to as the `forward direction', and the
$X$ axis pointing left towards the centre of HERA.  The coordinate
origin is at the nominal interaction point.  The laboratory
pseudorapidity, $\eta$, is defined as $-\ln\tan(\theta/2)$, where the
polar angle, $\theta$, is measured with respect to the proton beam
direction. All kinematical calculations take into account the position of the 
event vertex.}
 $36.7^{\circ} < \theta < 129.1^{\circ}$.  The electromagnetic cells
have a projective geometry as viewed from the interaction point.  Each
is 23.3 cm long in the azimuthal direction, representing 1/32 of the
full 360$^{\circ}$, and has a width of 4.9 cm along the beam direction
at its inner face, at a radius 123.2 cm from the beam line.  The
hadronic section consists of non-projective cells, each of which
covers four electromagnetic cells.  The azimuthal position of a
single-particle impact point within a cell is measured from the ratio
of the signals read out by photomultiplier tubes at each end, giving a
measurement with a resolution of $\pm2.5$ cm. The
photons were distinguished from neutral mesons ($\pi^0,\; \eta^0$) by
means of variables derived from the clusters of calorimeter cells
identified as electromagnetic signals, using the same method as
employed in previous ZEUS analyses~\cite{Z1,Z2}.  The most important
variable is the fraction of the cluster energy found in the cell 
with most energy, $f_{max}$, which peaks near unity for
signals from single photons.  After applying a cut to remove
candidates with a large cluster width, the events in any given bin of
a plotted physical quantity were divided into two classes with,
respectively, high and low values of $f_{max}$.  Information from
simulated single high-energy photons, $\pi^0$ mesons and $\eta^0$
mesons was then used to perform a statistical subtraction of the
background from the photon signal.

To reduce the backgrounds and the contribution from high-energy
photons radiated from outgoing quarks, an isolation criterion was
applied.  Within a cone of unit radius in pseudorapidity and azimuth
$(\eta, \phi)$ surrounding an outgoing photon candidate, the
integrated transverse energy in the detector, excluding that of the
photon candidate, was required not to exceed 10\% of that of the
photon candidate itself.  Both calorimeter cells and tracks were taken
into account in evaluating this condition.  In addition, no photon
candidate was permitted to have a track pointing within 0.3 radians of
it. Given the excellent performance of the ZEUS CTD, this effectively
assured that no electrons were misidentified as photons.

The trigger for the prompt photon events required an electromagnetic
energy cluster in the barrel section of the calorimeter, together with
further calorimeter requirements on the total energy of the event.
The offline cuts were at an adequate margin above the trigger level. 
In the offline analysis, use was made of energy-flow objects which
combine information from calorimeter cells and measured
tracks~\cite{zufos}.  For each event, the energy of the incoming
virtual photon was estimated using the quantity $\yJB$ = \mbox{$\sum(E
- p_Z)/2E_e$,} where the sum is over all energy-flow objects in the
event, each of which is treated as if due to a massless particle with
energy $E$ and longitudinal momentum component $p_Z$.  After
correcting for the effects of energy losses, limits of $0.20 < \yJB <
0.70 $ were applied, approximately corresponding to a centre-of-mass
$\gamma p$ energy range $134 < W < 251$ GeV.  The lower limit removed
proton beam-gas events, and the upper limit removed deep inelastic
scattering events.

Events with a scattered beam positron identified in the calorimeter
were rejected. The virtuality of the incoming photon was in this way
limited to values below $\approx 1$ GeV$^2$ with a median of $10^{-3}$
GeV$^2$.  Jets were reconstructed, using energy-flow objects, by means
of the Lorentz-invariant $k_T$-clustering algorithm KTCLUS
\cite{ktclus} in the inclusive mode~\cite{ellis}.  The standard
settings were used.  Corrections to the measured photon and jet
energies were evaluated through the use of Monte Carlo event samples
and were typically +5\% to +10 \% for both the photon and the jet.
After correction, photons were required to have $E_T^\gamma > 5$ GeV
and $-0.7 < \eta^\gamma < 0.9$, while jets were required to have
transverse energy $E_T^{\jet} > 5$ GeV, with pseudorapidity in the
range $-1.5 < \eta^{\jet} < 1.8$.  These kinematic cuts confined both
photons and jets to well-measured regions.  The momentum components of
the objects comprising the jet were summed to obtain the total
jet-momentum vector.  If more than one jet was found within the above
kinematic limits, the jet with highest $E_T$ was taken.  After the
above cuts and the cut on cluster width, the number of events with a
prompt photon candidate and a jet was 1507, of which approximately
half were background.

The fraction of the incoming photon energy that takes part in the QCD
subprocess was estimated by evaluating $x_\gamma^\meas$, defined
as~\cite{Z1}
$$x_\gamma^\meas = 
 \frac{1}{2E_e\,\yJB}\,\sum_{\gamma, {\jet}}\,(E-p_Z),$$
where the sum is over the high-energy photon and the contents of the
jet. The $x_\gamma^\meas$ distribution peaks at values close to unity
for direct photoproduction events, in which the whole photon takes
part in the hard subprocess.  It takes smaller values for resolved
events, where the photon acts as a source of partons, one of which
takes part in the hard subprocess.

\section{Results}

Figure 1(a) shows the distribution of $x_\gamma^\meas$ after the
subtraction of background due to $\pi^0$ and $\eta^0$ mesons.  The
errors shown are statistical only; systematic errors are dominated by
uncertainties on the parameters of the background subtraction and on
the calorimeter energy scale, and are typically $\pm7$\%.  It is
evident that both direct and resolved processes are present.  The
histograms show predictions from the PYTHIA 6.129 Monte
Carlo~\cite{Pythia}, after the events have been passed through a full
GEANT-based simulation~\cite{GEANT} of the ZEUS detector.  Default
settings of PYTHIA 6.129 were used, together with the parton density
functions MRSA~\cite{MRS} for the proton and GRV~\cite{GRV} for the
photon.  Approximately four times as many Monte Carlo events as data were
generated.  The PYTHIA distribution includes events from
direct and resolved prompt photon photoproduction at lowest order in
QCD, together with radiative dijet events in which an outgoing quark
from a hard QCD scatter radiates a high-energy photon that satisfies
the present experimental selections.  Figure 1(b) shows the
pseudorapidity distribution of the photons, the presence of a jet
being required.  The agreement with PYTHIA in both distributions is
qualitatively satisfactory, although the predictions lie below the
data, particularly at negative $\eta^\gamma$ values. This was also
observed in the ZEUS inclusive prompt photon measurements~\cite{Z2}.

Figure 2 shows distributions of kinematic quantities of the prompt
photon + jet system, for events selected with $x_\gamma^\meas > 0.9$.
These events are predominantly from direct photoproduction processes;
the restriction to high $x_\gamma^\meas$ also suppresses events with
additional jets.  Since the value of \kt\ affects primarily the shapes
of the distributions, these are area-normalised.  As illustrated in
Fig.\ 2(a), the plotted quantities describe the momentum imbalance of
the photon-jet system projected on to the transverse plane. These
quantities are the momentum component of the photon perpendicular to
the jet direction, $p_\perp$; the momentum imbalance along the
direction opposite to that of the jet, $p_\|$; and the azimuthal
acollinearity between the photon and the jet, $\Delta\phi$.  In
plotting $p_\|$, the condition $(p_T^{\jet} + p_T^\gamma) > 12.5$ GeV
was applied to remove an enhancement around zero due to the many
events where the photon and jet transverse energies both lie just
above their respective lower cuts.  The quantity $p_T^\jet$ is not the
Snowmass $E_T^{\jet}$, defined as the scalar sum of the $E_T$ values
of the individual particles in the jet, but is the transverse
component of the vector sum of the momenta of the particles in the
jet.  The quantity $\Delta\phi$ is strongly correlated with $p_\perp$
but has less sensitivity to the measured photon and jet energies.
%
%

To evaluate the systematic effects, the parameters of the background
subtraction and the calorimeter energy scale were varied within their
uncertainties.  For $p_{\perp} < 1$ GeV and $\Delta\phi > 170^{\circ}$, 
the effects were typically 1-2\% after normalisation of the
distributions.  Since there is some disagreement with theory at low
$\eta^\gamma$, the photon rapidity range was also reduced at each end
by 0.2. To observe the effects of including a greater proportion of
resolved events, the cut on $x_\gamma^\meas$ was reduced to 0.85.  As
further checks, the mode of jet reconstruction was varied by (a)
changing the jet-recombination scheme between the $p_T$ and $E$ modes
\cite{ktclus}, (b) using a cone jet algorithm~\cite{EUCELL}, and (c)
increasing the jet-radius parameter in the KTCLUS algorithm by a
factor 1.25.  The latter has sensitivity to the possible
jet-broadening effects of hard final-state gluon radiation.  The
results in each case were consistent with the main method at the level
of the statistical uncertainties.  The largest systematic
uncertainties come from the variations on the $x_\gamma^\meas$ and
$\eta^\gamma$ cuts. The effects of the incoming photon virtuality can
be neglected in the present analysis.  A relaxation of the 
lower cut on $\yJB$ did not significantly affect the result.

The data are compared to different predictions from PYTHIA, which
include the small contributions from radiative and resolved events
with $x_\gamma^\meas > 0.9$.  All the Monte Carlo results used here
are based on samples selected with the same detector-level cuts as the
data.  Within PYTHIA it is possible to vary the `intrinsic' smearing
on the transverse momentum of the partons in an incoming hadron; this
smearing is imposed in addition to the effects of parton showering,
and results are shown for three values of its two-dimensional Gaussian
width, \ks, as applied within the proton.  The corresponding mean
absolute value of the intrinsic transverse parton momentum in the proton,
\ktintr, is then given by $\ktintr = \sqrt{\pi/4}\,
\ks$~\cite{summary}.  In PYTHIA 6.129, the default value\footnote{In
later versions of PYTHIA, this default value has been increased to
approximately 1 GeV.}  
of \ks\ for partons in the proton and in the resolved photon is 0.44
GeV.  The photon \ks\ was fixed at this value.  The sensitivity of the
present measurement to the photon \ks\ is small since the selected
events are predominantly from direct processes.  From the figures it
is seen that the $p_\|$ distribution lacks discriminating power, which
prevents the use of the overall transverse
momentum, $Q_T$, of the photon - jet system to measure \kt, since this
depends substantially on $p_\|$.  It is evident that proton \ks\
values of $\ge$ 3 GeV or $\le$ 0.44 GeV are disfavoured by the
distributions in $p_\perp$ and $\Delta\phi$.

Normalised cross sections for $p_\perp$ and $\Delta\phi$ are presented
in Fig.\ 3.  The data have been corrected to the hadron level,
applying the same kinematic cuts as for the corrected detector-level
quantities, namely: $E_T^\gamma > 5$ GeV, $-0.7 < \eta^\gamma < 0.9$,
$E_T^{\jet} > 5$ GeV, $-1.5 < \eta^{\jet} < 1.8 $, $134 < W < 251$
GeV, and $x_\gamma^\meas>0.9$.  The PYTHIA prediction corresponding to 
a fitted value of \ks\ (see below) describes the data well.  The
systematic uncertainties are similar to those on the corresponding
distributions of Fig.\ 2 and are small compared to the statistical
uncertainties.  Uncertainties in the calorimeter energy scale largely
cancel in the normalised  distributions.  Within PYTHIA, the 
r.m.s.\ width in $\phi$ between the directions of the outgoing parton 
and the reconstructed jet was found to be $\pm6.4^\circ$, considerably
less than the corresponding value of $\pm12.9^\circ$ between the 
jet and the reverse of the photon direction.

Further PYTHIA Monte Carlo samples were generated using proton \ks\
values of 1.0 GeV and 2.0 GeV, in addition to those shown in Fig.\ 2.
A $\chi^2$ minimisation was performed to determine the optimal value
of \ks\ using the $p_\perp$ data and the five PYTHIA simulations at
the detector level.  The resulting fitted \ktintr\ value is
$1.25\pm0.41\; ^{+0.15}_{-0.28}$ GeV, where the first error is
statistical and the second systematic.  A fit to the $\Delta\phi$
data, which is highly correlated with $p_\perp$, gave a similar result
with a larger statistical uncertainty.  The value of \ktintr\ does not
include the parton-shower contribution to \kt, which was found to be
approximately 1.4 GeV from a PYTHIA sample with \ks\ = 0.  Within the
framework of direct photoproduction in PYTHIA, the contributions to
$p_\perp$ at the parton level from the intrinsic component and from
the parton shower may be combined to allow a total value
of \kt\ to be evaluated assuming that the overall distribution is
Gaussian\footnote{Assuming 
Gaussian distributions, the relationship $\kt = \sqrt{\pi/2}
\pprms$ holds.  By varying \ks\ in PYTHIA and noting the resulting
distributions of \pprms, \kt\ and \ktintr\ at the parton level, for
events passing the experimental cuts at the detector level, the
relationship $\kt^2 = 1.92\mbox{~GeV}^2 + 0.61\ktintr^2$ was
obtained.}.
The total \kt\ value corresponding to the fitted value of \ktintr\ was
evaluated in this way to be
$$ \kt = 1.69\pm0.18\; ^{+0.18}_{-0.20} \mbox{ GeV}. $$ 
The systematic error includes a contribution from the model-dependence
of the result, as estimated from a calculation using HERWIG 6.1
\cite{HERWIG}.  HERWIG uses a parton-showering model which differs
from that used in PYTHIA; in particular, it does not have the sharp
lower cut-off in the shower evolution below which PYTHIA relies upon a
suitable value of its phenomeno\-logical \ks\ parameter~\cite{TB}.  It
was found that the data were already well described by HERWIG with its 
\ks\ at the default value of zero, so that the fitting method used above
could not be repeated.  The use of the default version of HERWIG gave
a \kt\ value 10\% higher than the PYTHIA result; this was added to the
systematic error as a contribution of $\pm$ 10\%.

Both PYTHIA and HERWIG model the effects of final-state gluon radiation.
As a further check, a sample of PYTHIA events was prepared with the
final-state radiation turned off.  The results were not significantly
different from those using the standard version.  

The value of \kt\ determined here represents the mean absolute value of the
parton transverse momentum in the proton, taking into account all the
partonic effects modelled in direct photoproduction within PYTHIA.
It may be compared with values of \kt\ obtained previously by other
methods, such as by the direct measurement of outgoing lepton or
photon pairs.

Figure 4 shows the ZEUS result in comparison with mean \kt\ values from other
experiments~\cite{summary,MB}, plotted as a function of the hadronic
centre-of-mass energy, $W$, of the incoming particles, namely the
photon and proton in the present case.  At low $W$, the data come
mainly from the production of muon pairs by the Drell-Yan mechanism in
fixed-target interactions, while the production of photon pairs has
been studied at high $W$ using the ISR~\cite{ISR} and Tevatron colliders.  The
ZEUS result bridges a gap between the low- and high-energy measurements.
For a uniform presentation, the mean transverse momentum, \Qt, 
frequently quoted in the production of photon and muon pairs by pairs
of incoming hadrons has, where necessary, been converted to a \kt\ value for a
single hadron by dividing by $\sqrt{2}$.  It should be noted that
while the measurement of final-state dimuon and photon pairs provides
a direct determination of parton \kt, the other measurements require a
physical model.  Although different experimental methods have been
employed, a clear trend for \kt\ to rise with increasing $W$ is
evident, as discussed most recently by Laenen et al.~\cite{gluons2}.
The ZEUS result is fully consistent with this trend.

\section{Conclusions}

Photoproduction events containing a prompt photon balanced by a recoil
jet have been studied using the ZEUS detector at HERA.  Events in the
$\gamma p$ centre-of-mass energy range $134 < W < 251$ GeV were
selected containing a photon with $E_T^\gamma > 5$ GeV and $-0.7 <
\eta^\gamma < 0.9$ and a jet with $E_T^{\jet} > 5$ GeV and $-1.5 <
\eta^{\jet} < 1.8$.  The kinematic properties of the photon-jet system
were used to investigate the effective transverse momentum of the
quarks in the proton, within the framework of the PYTHIA 6.129 Monte
Carlo.  A fit to the data gave a \kt\ value of $1.69\pm0.18\;
^{+0.18}_{-0.20}$ GeV.  This result is consistent with the trend,
observed in a number of experiments at different energies, that the
effective parton \kt\ rises with the energy of the interacting
hadronic system.

\section*{Acknowledgements\\[-13mm]}
As always, it is a pleasure to thank the DESY directorate and staff for
their support and encouragement.  The outstanding efforts of the HERA
machine group are likewise gratefully acknowledged, as are the many
technical contributions from members of the ZEUS institutions who are
not listed as authors.  We thank M. Begel, M. Seymour and
T. Sj\"ostrand for helpful correspondence.\\

\newpage
\begin{figure}\centerline{
\epsfig{file=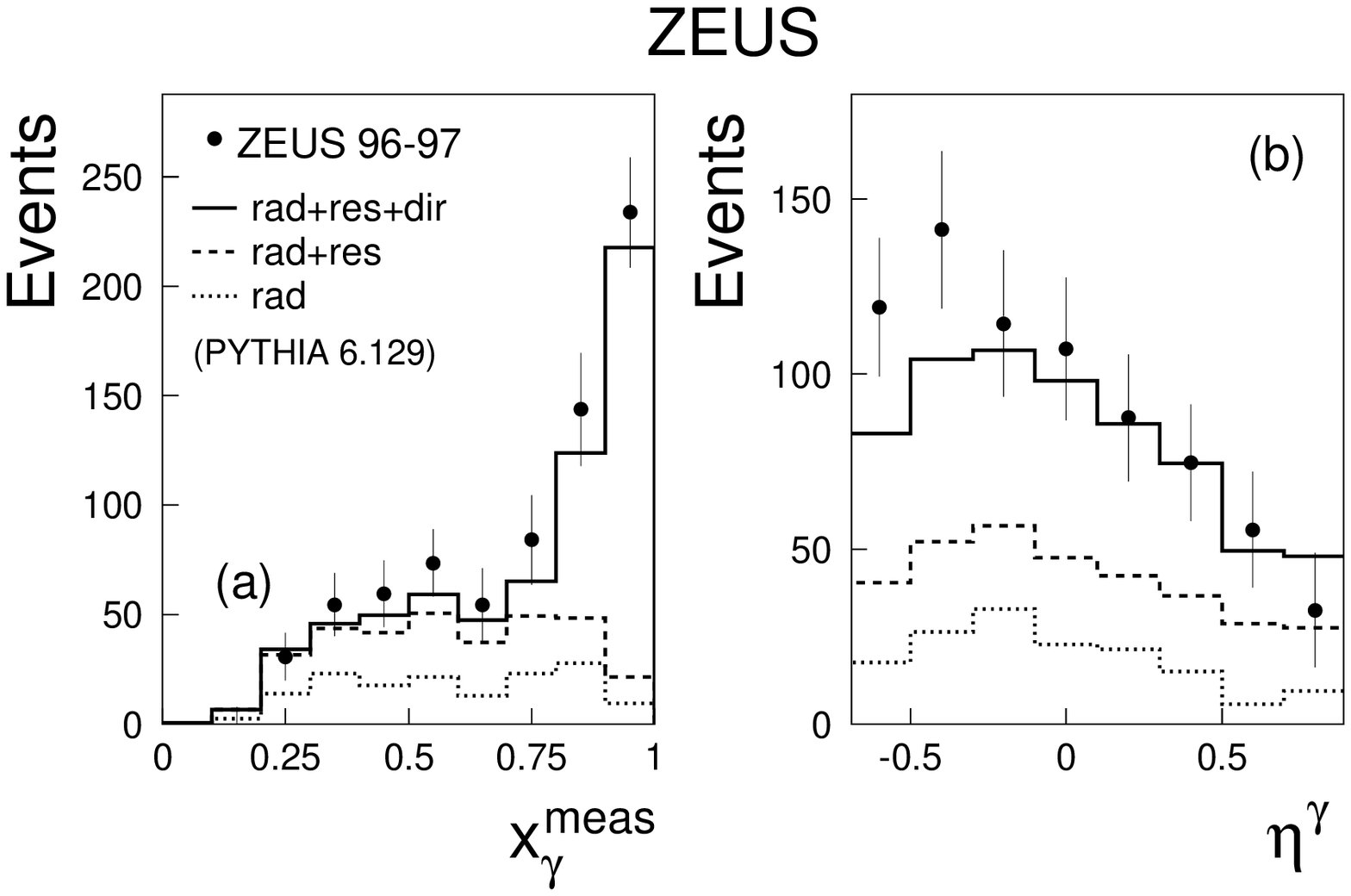,width=15cm}
}
\caption{Distributions of (a) $x_\gamma^\meas$, (b) photon
pseudorapidity, $\eta^\gamma$, for prompt photon events in which a jet
is also observed. The plotted errors are statistical only. The data
are compared with predictions from PYTHIA 6.129.  The PYTHIA
histograms indicate contributions from dijet events where a
final-state quark radiates a photon (dotted line), summed with
resolved prompt photon events (dashed line), and also with direct
prompt photon events (full line).  The PYTHIA 6.129 default \kt values
in the proton and photon have been used. The Monte Carlo predictions are 
normalised to the integrated luminosity of the data.}
\end{figure}

\newpage
\vspace*{-1cm}
\begin{figure}\centerline{
\epsfig{file=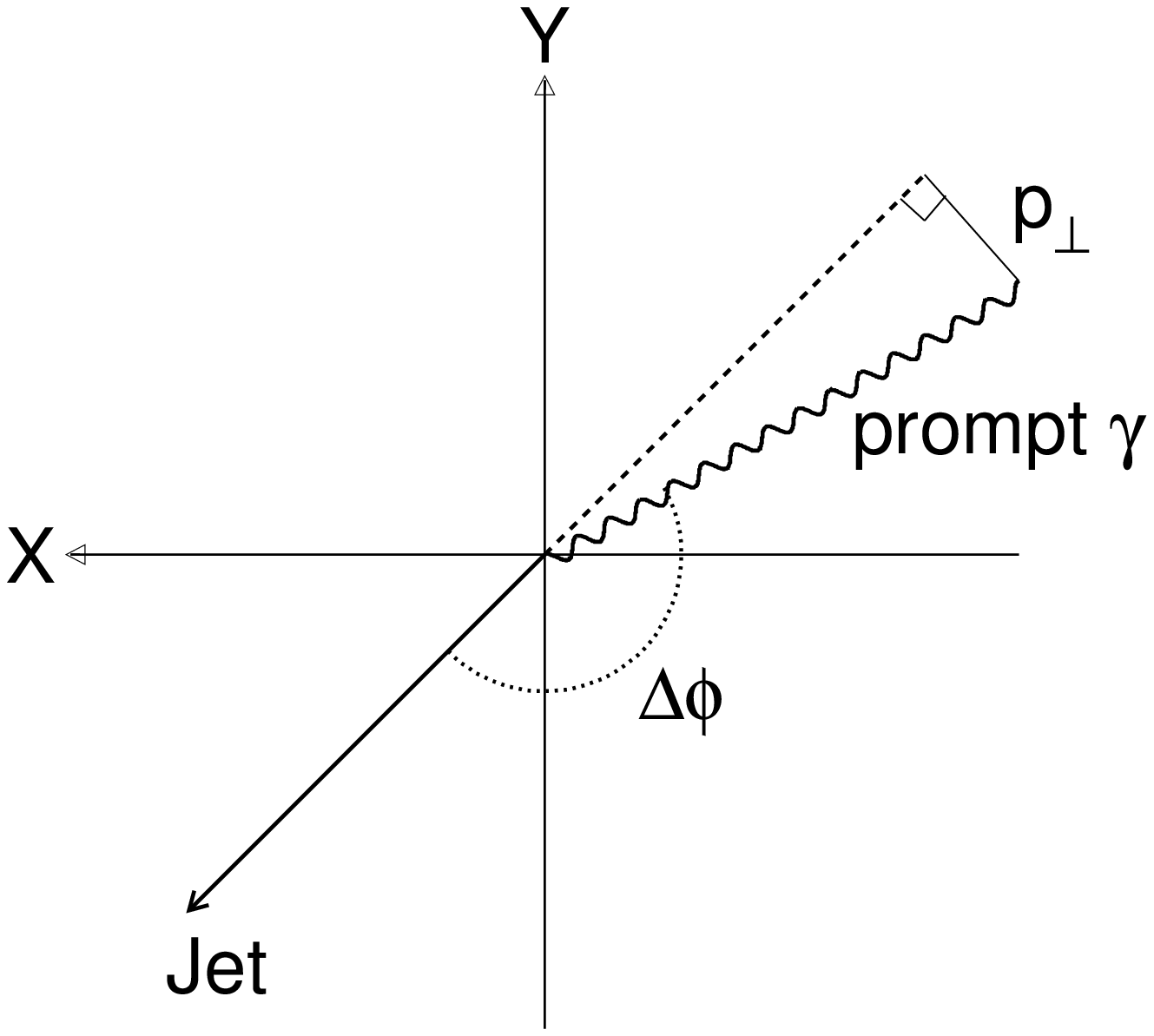,width=9cm}\hspace*{-8.5cm}
\raisebox{-9cm}{
\epsfig{file=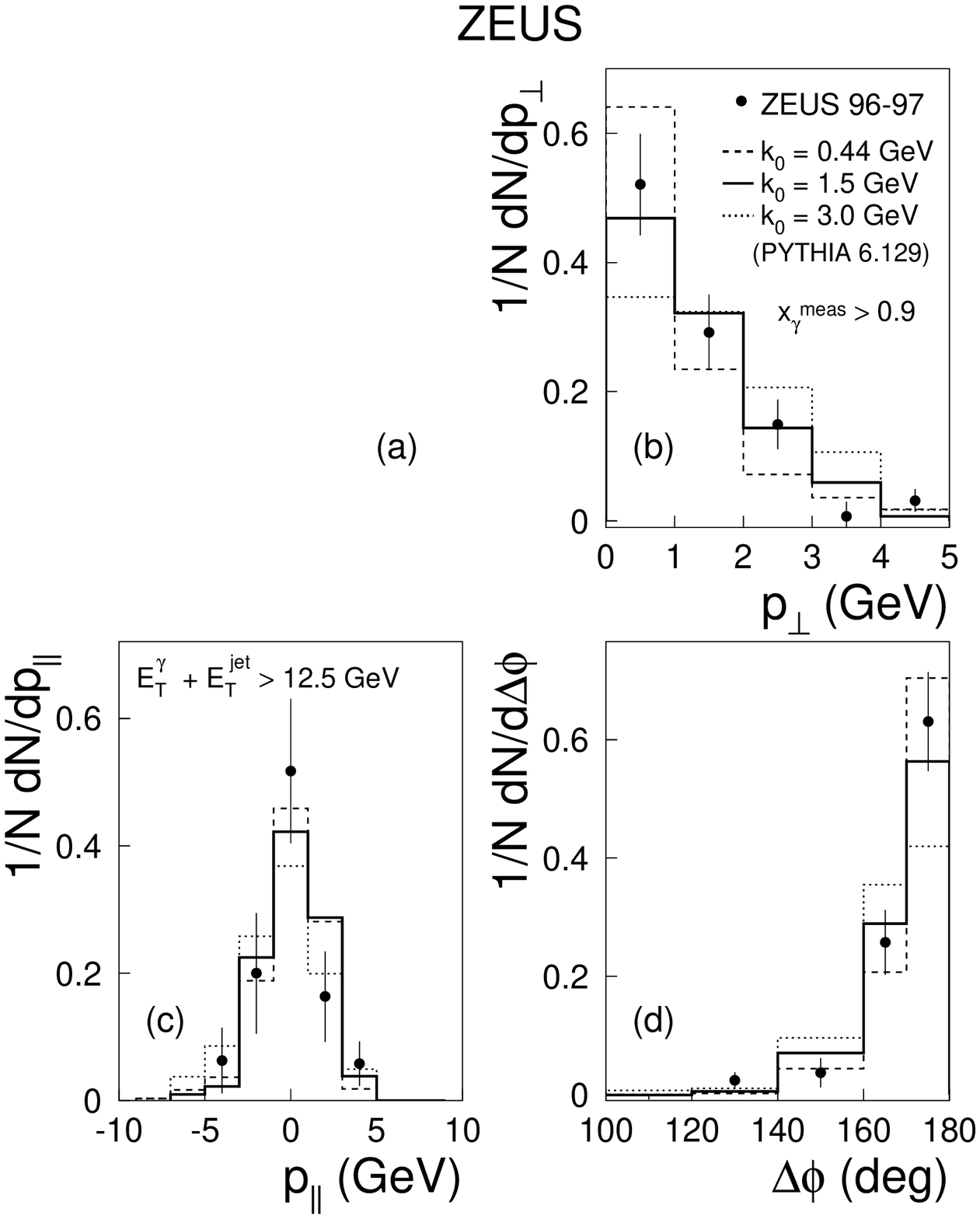,height=18.5cm}
}\hspace*{1cm} 
}
\vspace*{5mm}
\caption{Normalised detector-level distributions of kinematic
quantities observed in the production of a prompt photon with a jet,
compared with predictions from PYTHIA 6.129 generated with different
values of the `intrinsic' transverse momentum, \ks, of the partons in
the proton.  Only events with $x_\gamma^\meas > 0.9$ are used.  In (a)
the configuration of the photon and jet in the plane transverse to the
beam direction is illustrated.  The plotted quantities, calculated in
this plane, are: (b) perpendicular momentum component of the photon
relative to the jet direction; (c) longitudinal momentum imbalance
(photon -- jet) along the jet direction; (d) difference in azimuthal
angle between the photon and jet directions.  Statistical errors only
are shown.}
\end{figure}
\newpage
\begin{figure}\centerline{
\epsfig{file=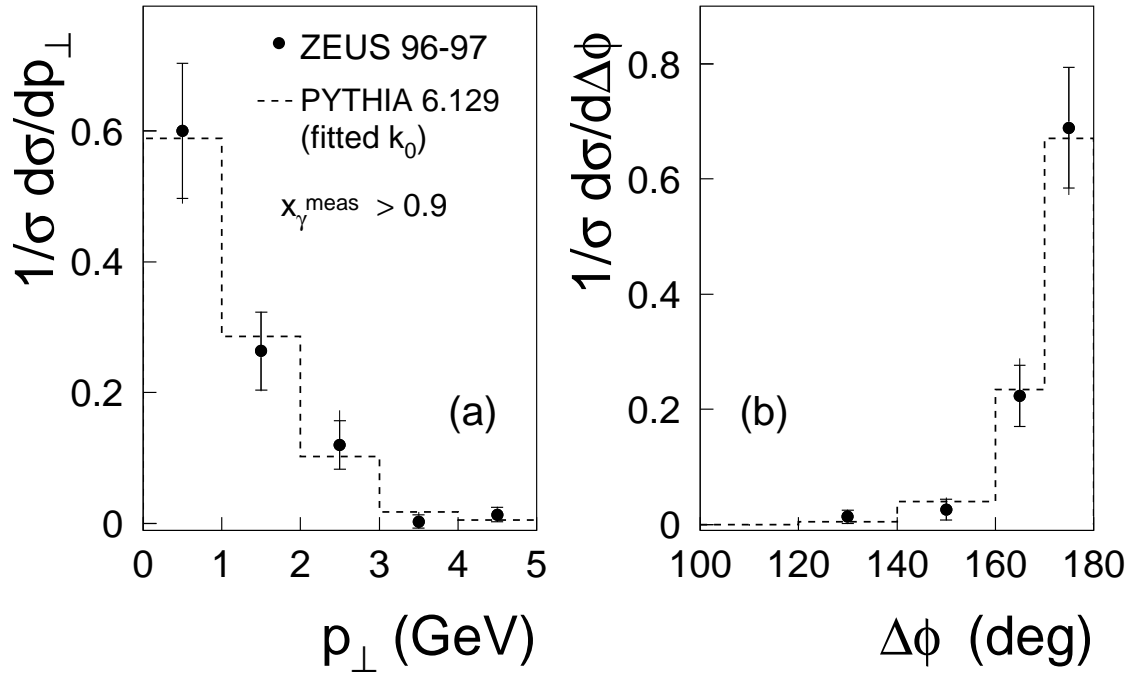,width=15cm,%
bbllx=30pt,bblly=380pt,bburx=520pt,bbury=710pt,clip=yes}
\vspace*{10mm}
}     

\caption{Normalised cross sections of kinematic quantities observed in
the production of a prompt photon with a jet, compared with
predictions from PYTHIA 6.129 corresponding to a fitted value of \ks\
= 1.42 GeV.  The inner and outer error bars represent statistical and
total uncertainties, respectively.  Only events with $x_\gamma^\meas >
0.9$ are used. The plotted quantities, calculated as in Fig.\ 2 but at
the hadron level, are: (a) perpendicular momentum component of the
photon relative to the jet direction; (b) difference in azimuthal
angle between the photon and jet directions.  }
\end{figure}

\newpage
\begin{figure}\centerline{
\epsfig{file=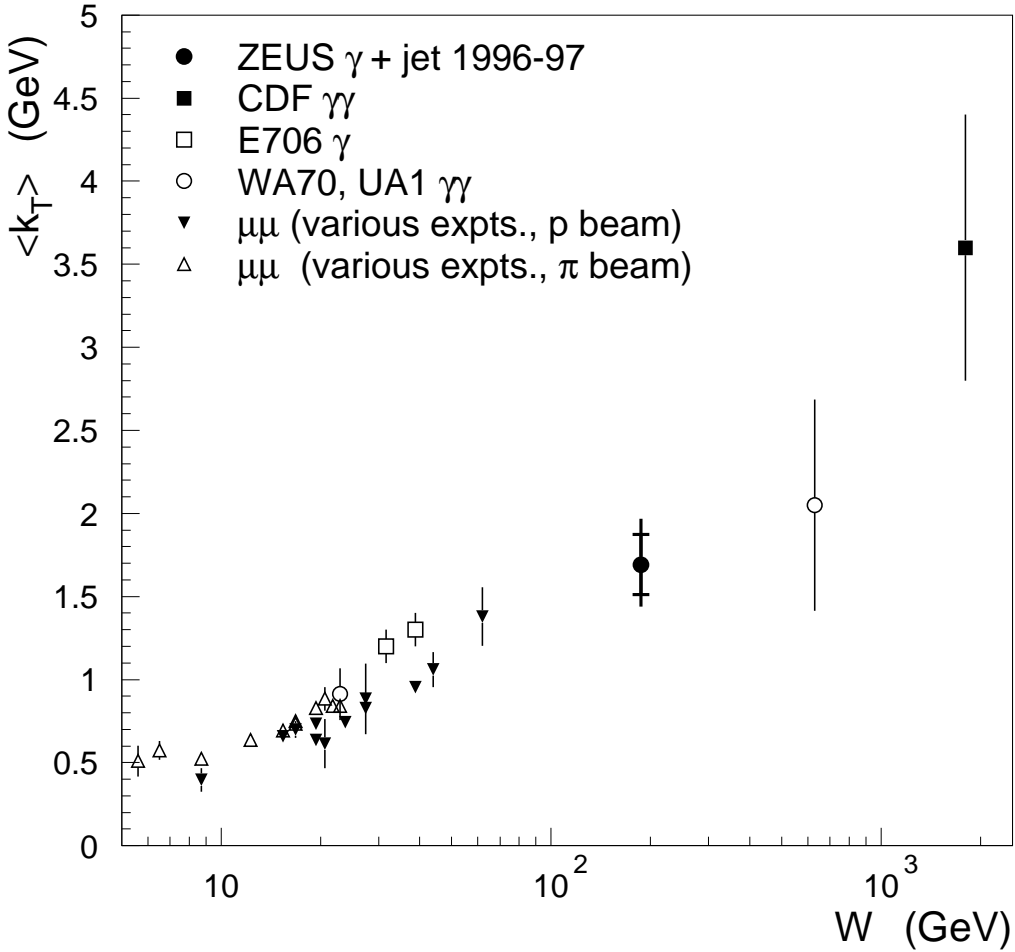,width=15cm}
}
\caption{The ZEUS measurement for \kt\ compared with results from
other experiments.  The inner and outer error bars on the ZEUS point
represent statistical and total uncertainties, respectively.  Other
published results have been scaled by $\sqrt{2}$ as appropriate (see
text).  The single prompt photon results from CDF and D0 are in
agreement with the double prompt photon CDF
data-point~\protect\cite{summary}. Full references~\cite{MB} may be
found in a recent FNAL report~\protect\cite{wgreport}.  The horizontal
axis denotes the centre-of-mass energy of the colliding particles,
which in the case of ZEUS are the photon and proton.}
\end{figure}

\end{document}